\documentclass[11pt]{amsart}
\usepackage{fullpage}
\usepackage[utf8]{inputenc}
\usepackage[T1]{fontenc}
\usepackage{natbib}
\usepackage[foot]{amsaddr}

\usepackage{amsmath,amsfonts,amsthm} % Math packages
\usepackage{graphicx}
\usepackage{subfig}
\graphicspath{{images/}}
\begin{document}

%\preprint{}

%\title{Extreme wave statistics in combined windsea and swell} %Title of paper
\title{Extreme wave statistics in co-propagating windsea and swell} %Title of paper



\author{Susanne St{\o}le-Hentschel$^\dagger$}
\address{$^\dagger$Department of Mathematics, University of Oslo, 0316 Oslo, Norway}
\date{\today}
\author{Karsten Trulsen$^\dagger$}
\email[Corresponding author: ]{karstent@math.uio.no}
\author{Shkurte Olluri$^\dagger$}

%\thanks{}
%\altaffiliation{Department of Mathematics, University of Oslo, Norway}
%\affiliation{Department of Mathematics, University of Oslo, 0316 Oslo, Norway}


%\date{\today}

\begin{abstract}
	We investigate how the extreme wave statistics of a windsea is modified by a following swell, by means of laboratory experiments and simulations using a High Order Spectral Method (HOSM) of long-crested sea.  The windsea spectrum is kept equal for all cases, while the swell is altered.  Analysis of the combined wave system gives the impression that the mixed sea has milder extreme wave statistics than the windsea alone, especially when the nonlinearities in the two systems clearly differ.  Upon partitioning the mixed sea into windsea and swell, the windsea part is found to be nearly unaffected by the swell and to be governed by essentially the same statistics irrespective of a swell being present or not.  This result is found for skewness, kurtosis and exceedance probability of envelope and crest height.
\end{abstract}

\maketitle
\section{Introduction}
Multiple wave systems constitute 15--25\% of sea states observed in different locations around the world ~\citep{GuedesSoares-1984-OE-11-185,GuedesSoares-1991-OE-18-167,Boukhanovsky-2009-AOR-31-132}. They may be distinguished by their interaction angle that is in the range of $[330^\circ,30^\circ]$ for following seas, $[150^\circ,210^\circ]$ for opposing seas and $[30^\circ,120^\circ]$ and $[240^\circ,300^\circ]$ for crossing seas~\citep{Toffoli-2005-AOR-27-281}. The analysis of 270 ship accidents, performed by ~\citet{Toffoli-2005-AOR-27-281}, showed that most of them occurred in crossing seas and were associated with a rapid change of the sea state. Among the crossing sea accidents analysed in more detail are the Louis Majesty incident ~\citep{Cavaleri-2012-JGR-117-C00J10} that occurred at a crossing angle of $40^\circ$--$50^\circ$ and the Prestige accident~\citep{Trulsen-2015-JGRO-120-7113} at a crossing angle of $90^\circ$. Crossing seas were also claimed as an important condition preceding the Suwa-Maru incident~\citep{Tamura-2009-GRL-36-L01607}. The latter happened in an 'extremely narrow'~\citep{Tamura-2009-GRL-36-L01607} unimodal swell that had developed from two swells and one windsea.

The hypothesis of increased freak wave probability due to mutual enhancement of two coexisting systems for crossing seas was also studied by ~\citet{Onorato-2006-PRL-96-014503} and ~\citet{Toffoli-2011-GRL-38-L06605}. 
~\citet{Onorato-2006-PRL-96-014503} investigated crossing seas theoretically by coupled nonlinear Schr\"odinger (CNLS) equations and found larger regions of instability and large growth rate for coupled systems
compared to single systems. Simulations of the CNLS were then compared with simulations of the higher order spectral method (HOSM) and laboratory measurements. The kurtosis was found
to be increased for two systems with an interaction angle between $40^\circ$  and $60^\circ$ ~\citep{Toffoli-2011-GRL-38-L06605}, indicating enhanced probability of freak waves according to ~\citet{Mori-2006-JPO-36-1471}.

Among the possible interaction scenarios, we focus on those targeting the interaction of windsea and swell, distinguished by two opposite interaction regimes. For interactions over a long scale, there is evidence that a windsea may modify a swell \citep{Masson-1993-JPO-23-1249} and may enhance the occurrence of freak swell waves \citep{Tamura-2009-GRL-36-L01607} through resonant transfer of energy from the windsea to the swell. This may happen when the separation in wave periods between the two systems is small. However, for the opposite regime of short interaction scale, the swell can modulate the amplitude and phase of the short waves \citep{Gramstad-2010-JFM-650-57}. They assumed that the nonlinear energy transfer between the two systems was limited by sufficient separation of the periods of the long and the short waves. The focus herein is to investigate whether such a modulation of the nonlinear wave evolution can lead to changes in the statistical properties of the wind waves and in particular if it makes freak waves more or less probable.

Thus far, it has been found that the a swell may modulate the windsea waves in such a way that the freak wave probability is increased.
\citet{Regev-2008-PF-20-112102} found that a windsea and a weak swell at right angle  can suffer wave modulations that lead
to freak waves. Similarly, \citet{Gramstad-2010-JFM-650-57} computed the modification of the probability of freak waves in a windsea perturbed by a weak swell oriented at various angles to the windsea. They found that the swell could enhance the occurrence of freak waves in the windsea slightly. However, in the case of right angle between the swell and the windsea the increase was found to be minimal.
A different path from swell and windsea interaction to freak wave generation was suggested by \citet{Tamura-2009-GRL-36-L01607} who speculated that the nonlinear coupling between swell and windsea could generate a narrow spectrum. Some ship accidents indeed seem to have occurred in conditions of narrowing wave spectra, e.g.\ \citet{Tamura-2009-GRL-36-L01607, Waseda-2012-JMST-17-305, Waseda-2014-NHESS-14-945}.

Following \citet{LonguetHiggins-1960-JFM-8-565}, a large number of studies was devoted to how longer waves modulate shorter waves, e.g.~\citet{LonguetHiggins-1987-JFM-177-293,Craik-1988-JAMSB-29-430, Grimshaw-1988-JAMSB-29-410, Henyey-1988-JFM-189-443, Zhang-1990-JFM-214-321, Zhang-1992-JFM-243-51}. 
The amplitudes of the short waves are enlarged and their wavelength is shortened close to the crests of the underlying long waves. Near the troughs of the long waves, the short waves obtain longer wavelengths and smaller amplitudes. \citet{Zhang-1992-JFM-243-51} concluded that the modulational instability of weakly nonlinear short waves was enhanced when they were riding on finite-amplitude long waves. Similar results for the nonlinear evolution of short gravity waves on long waves were also obtained by \citet{Naciri-1992-JFM-235-415, Naciri-1994-WM-20-211,Naciri-1993-PFA-5-1869}. First they studied short gravity waves on long rotational Gerstner waves \citep{Naciri-1992-JFM-235-415}, then irrotational short waves on irrotational long waves \citep{Naciri-1993-PFA-5-1869} and then two-dimensional interaction of obliquely intersecting waves \citep{Naciri-1994-WM-20-211}. For the latter, the instability of the short wave due to oblique side bands was shown to be enhanced by the presence of the long wave. However, the obliqueness becomes important only when the steepness of the long wave is sufficiently large.

The work presented herein concerns the wave statistics of different combinations of swell and windsea. Mixed seas have earlier been studied by \citet{petrova2009probability,petrova2011wave,Petrova2014distributions}, with the main focus on wave height distributions.
%KT that will not be discussed here due to their ambiguous definition.
Their laboratory experiments investigate mixed seas of equal significant wave height. For swell dominated seas, they found that skewness and kurtosis are lower than in windsea dominated seas \citep{Petrova2014distributions,petrova2009probability}. They also found that the exceedance probability of windsea dominated cases have thicker tails than those of swell dominated cases \citep{petrova2009probability,petrova2011wave,Petrova2014distributions}. 
We believe these observations may be explained by the fact that the portion of the steeper windsea waves decreases and the portion of swell waves with low steepness increases. Our experimental setup is different from theirs in that the windsea contributions are kept the same irrespective of the strength of the swell.

%The results are presented in six sections: Measured kurtosis and skewness for different sea states;  Simulated kurtosis and skewness for different sea states; Exceedance probability of wave envelope; 
%     Exceedance probability of crest height; Spectral evolution along the tank; Statisitcal analysis revisited for split spectra.

%KT
% We anticipate that the presented work has general implications for wave analysis of sea states with multiple peaks.

%     Remarks to this effect may be found in the discussion and the conclusion.

%KT
First, we summarize briefly our setup in the laboratory and simulations and define the characteristics of the analyzed sea states.
Then, we compare the analysis of skewness and kurtosis for the partitioned windsea and the mixed seas. 
Finally, we investigate the exceedance probabilies for the partitioned and combined cases.
Our main conclusion is that in order to fully comprehend the combined extreme wave statistics, mixed seas should be partitioned, and analysis be performed on each partition.

%\ifnum1=0

%----------------------------------------------------------------------------------------
\section{Setup of the laboratory experiments}
%----------------------------------------------------------------------------------------
\begin{figure}
	\centering
	\includegraphics[width=0.8\textwidth]{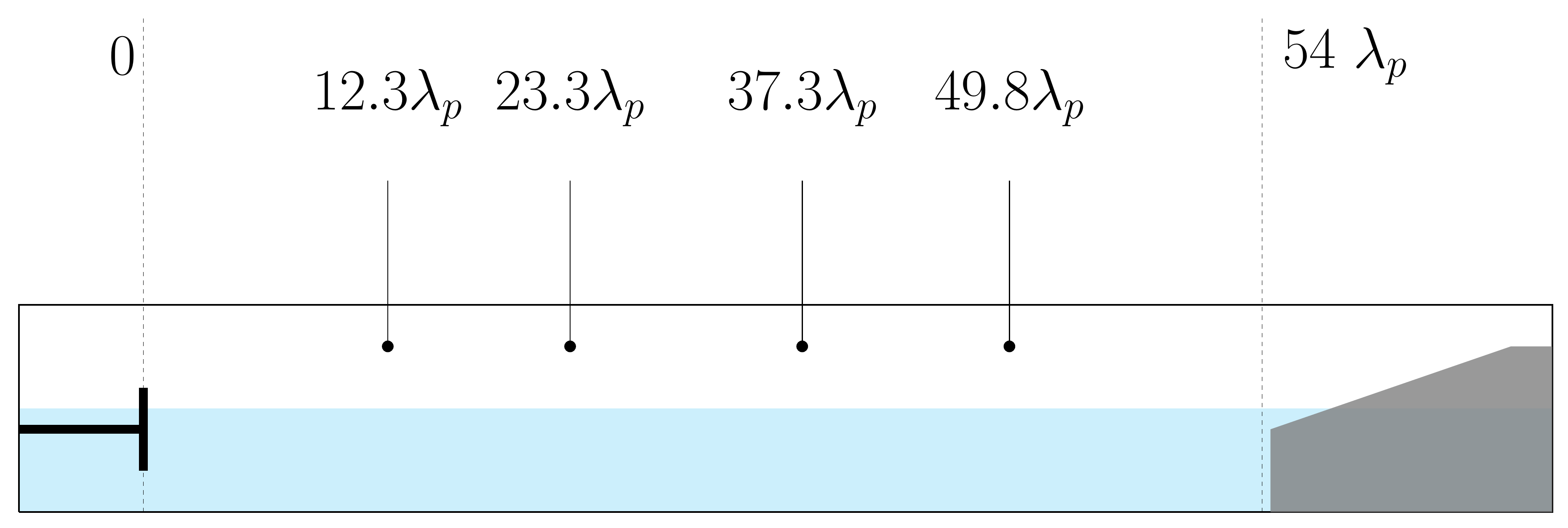}
	\caption{Sketch of laboratory experiments in the wave tank. The wave maker is placed on the left, the 
		damping beach on the right and the four ultrasonic probes marked by their distances from the wavemaker in terms of the peak wavelength $\lambda_p$.}
	\label{fig:wave_tank}
\end{figure}

The experiments were performed in a wave flume at the Department of
Mathematics at the University of Oslo. The dimensions of the tank and the
setup are sketched in Figure\ \ref{fig:wave_tank}. The wave tank is 24.6 m
long, 0.5 m wide and was filled to a mean water depth $h=0.7$ m.

Unidirectional waves were mechanically generated at one end of the tank by a hydraulic
piston, programmed according to a Torsethaugen spectrum with characteristics described below.

A 3 m long beach at the opposite end of the tank damped the waves to prevent reflection. 
The surface elevation was measured with four ultrasonic probes at
the positions indicated in Figure~\ref{fig:wave_tank}. Key parameters for the laboratory experiment are listed in Tab.~\ref{tab:sim_pars}.
Visual inspection during the
experiments did not reveal wave breaking, although the measurements do indicate dissipation.

\begin{table}
	\caption{Key parameters for laboratory experiments and simulations.}
	\label{tab:sim_pars}
	%\large
	\normalsize
	\begin{tabular}{l c c   }
		& Laboratory & Simulation\\
		\hline\hline
		Number of probes/points along the tank & 4 & 1024 \\
		Total time [$T_p$]  & 3058 & 277\\
		Start time for analysis  [$T_p$]  & 117 & 119 \\
		Analysis time [$T_p$] & 2822 & 158\\
		Number of simulations & 1 & 54\\
	\end{tabular}
\end{table}

The Torsethaugen spectrum applied herein is given 
in its simplified version for windsea dominated spectra \citep{Veritas2014}, 
\begin{align}
\begin{split}
&S(f) = \sum_{j=W,S} \frac{1}{16} H_{s_j}^2 T_{p_j}S_{n_j}(f_{n_j})\\
&S_{n_j} = 3.26 A_{\nu_j} \Gamma_{s_j}\nu_j\\
& A_{\nu_W} = \left[ 1+1.1\left( \ln(\nu) \right)^{1.19} \right]/\nu\\
& A_{\nu_S} = 1\\
&\Gamma_{s_j} = f_{n_j}^{-4} \exp\left[ -f_{n_j}^{-4} \right]\\
&f_{n_j} = f T_{p_j}\\
&\nu_W = \nu^{\exp\left[ -\frac{1}{2\alpha^2} \left( f_{n_W}-1 \right)^2 \right]}\\
&\nu = 35\left[\frac{2\pi H_{s_W}}{gT_{p_W}^2}\right]^{0.857}\\
&\alpha =\left\{\begin{array}{cc}
0.07 & f_{n_j}<1\\
0.09 & f_{n_j}\geq 1
\end{array}\right. \\
&\nu_S = 1
\end{split}
\end{align}
The values for the peak period, $T_p$, peak frequency, $\omega_p$, and significant wave height, $H_s$, are summarized in Table~\ref{tab:wave_pars}.
Parameters for windsea and swell are distinguished by the corresponding subindices $W$ and $S$. We have renamed the shape factor $\gamma$ in \citet{Veritas2014}
to $\nu$ to avoid collision with the skewness $\gamma$ to be discussed later. Similarly, we use $\alpha$ instead of $\sigma$, since the standard deviation $\sigma$ will play a role later in the document.

We believe that the peak wavelength is still long enough to suppress cross-tank modulations provoked by modulational instability, the parameter $\mu$ of \citet{Trulsen-1999-PF-11-235}
having the value $\mu=0.4\gg0.1$ for the peak period which should ensure we are outside the domain of transversal modulationally unstable modes.

%-------------------------------------------------------------------------------
\section{Setup of the numerical wave tank}
%-------------------------------------------------------------------------------

Numerical simulations were carried out with a higher order spectral method,
the HOS-NWT numerical wave tank described by \citet{Bonnefoy-2009-ANSNWW-11-129} and
\citet{Ducrozet-2012-EJMB-34-19}. We applied a 1D wave tank with the same scale as in the laboratory, spatially discretized at 1024 points. The waterdepth was set to five meters.
%The length scale was doubled in order to simulate with some distance to the cut-off frequency. (???where did I read that??)
A numerical wave paddle at one end of the tank was programmed to generate a two-peak spectrum 
\begin{align}
S(f) = \sum_{j={\mathrm{W,S}}}{  \frac{ H_{sj}^2 T_{pj} \left( \nu_j + 0.25 \right)^{\nu_j}  }{4 \Gamma(\nu_j) \left( T_{pj}f  \right)^{(4\nu_j + 1)}} \exp\left(- \frac{\nu_j + 0.25}{\left(T_p f\right)^4} \right) }
\end{align}
according to \citet{OchiHubble1976}.
The formulation is a superpostion of the frequency spectra of windsea, indexed by $\mathrm{W}$ and a swell spectrum, marked by the index $\mathrm{S}$. Both spectra are defined by the same formula depending on the significant wave height, $H_{s_j}$, the peak period, $T_{p_j}$, the shape factor $\nu_j$ which was set to 3.3 for swell and windsea and the Gamma function $\Gamma$.

The Ochi~\&~Hubble spectrum was preferred for the simulations since increasing the energy in the swell does not affect the energy content in the  windsea part as much as in the Torsethaugen model.
In addition, the swell was varied over a larger range in comparison to the laboratory, to capture a greater variation of cases.
The performance characteristics of the paddle were set to the default values. The
damping beach was simulated by the inherent numerical beach with an absorption
coefficient of unity~\citep{Bonnefoy-2009-ANSNWW-11-129,Ducrozet-2012-EJMB-34-19}.

Key parameters for the numerical simulations are listed in Tab.\
\ref{tab:sim_pars}.

\section{Characterization of the sea states}
\begin{figure}
	\centering
	\subfloat[][Laboratory]{\includegraphics[width=0.48\textwidth]{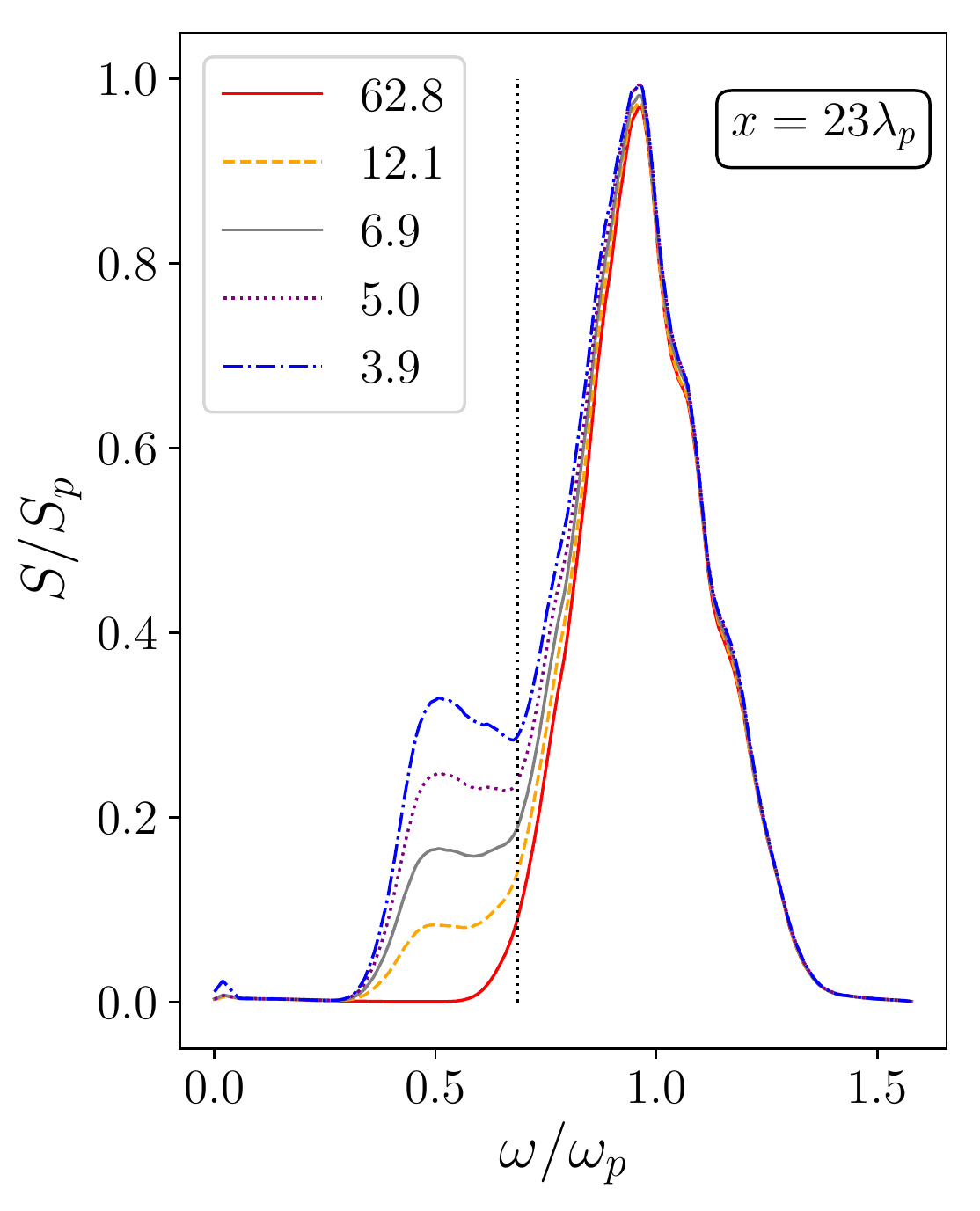}}
	\subfloat[][Simulation]{\includegraphics[width=0.48\textwidth]{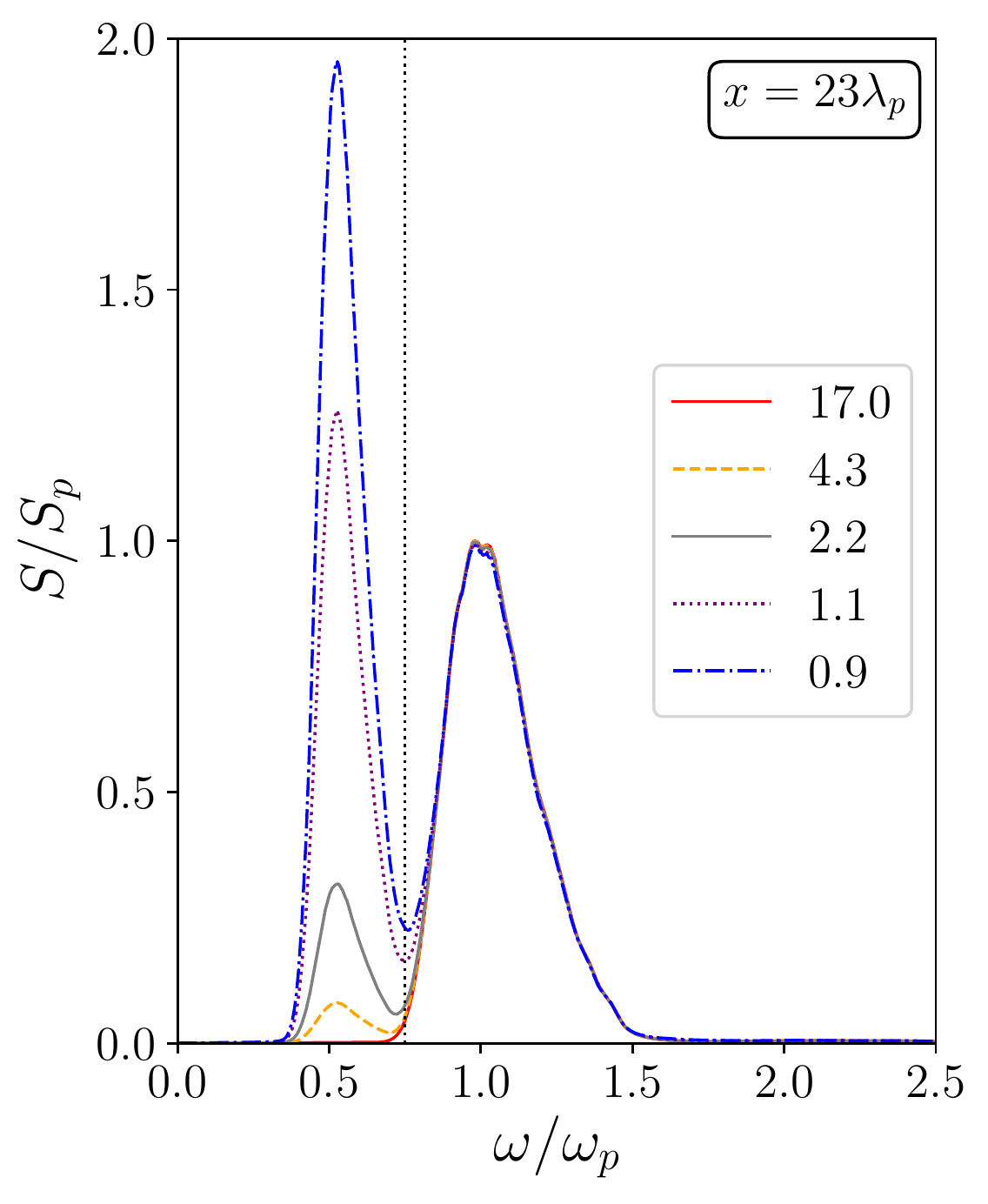}}
	\caption{Scaled spectra for the different laboratory experiments and simulations. The spectra are distinguished by the Sea Swell Energy Ratio (SSER) as defined by~\citet{GuedesSoares-1984-OE-11-185}.}
	\label{fig:spectra}
\end{figure}

%     KT dette er allerede sagt, ikke si det igjen
%     For the laboratory, the wave paddle was programmed to generate waves based on Torsethaugen, while Ochi \& Hubble spectra were employed in the simulations.
In the simulations the spectra transform fast in the vicinity of the wavemaker but beyond five peak wavelengths from the wavemaker the spectra retain their shape approximately. Figure~\ref{fig:spectra} shows an example of the spectra at the position $x=23~\lambda_p$. The vertical dotted lines mark the splitting frequencies $\omega_m$ that were employed to separate the combined swell and windsea ($\mathrm{SW}$) into swell ($\mathrm{S}$) and windsea ($\mathrm{W}$) contributions. Torsethaugen and Ochi \& Hubble spectra require different splitting frequencies, resulting in $\omega_m=0.685~\omega_p$ for laboratory and $\omega_m=0.75~\omega_p$ for simulation. The different sea states are labeled by the sea-swell energy ratio (SSER) as suggested by \citet{GuedesSoares-1984-OE-11-185},
\begin{align}
\mathrm{SSER} = \frac{m_{0\mathrm{W}}}{m_{0\mathrm{S}}},
\end{align}
based on the zeroth spectral moments   for windsea, $m_{0\mathrm{W}}$, and swell $m_{0\mathrm{S}}$.

One of the key techniques employed in the analysis below is the partitioning
of the two wave systems. The windsea was
extracted by highpass filter (HP) and the swell by lowpass filter (LP). 
A similar technique was employed by e.g. \citet{Trulsen-2015-JGRO-120-7113} and \citet{StoleHentschel-2018-PF-30-067102}.

Since the windsea spectrum is equal for all studied cases, its spectral peak $S_p=S_{p\mathrm{W}}$ at the frequency $\omega_p=\omega_{p\mathrm{W}}$ serves as scaling parameters and the indices for windsea are dropped for these two parameters.

The wave steepness and the Benjamin-Feir index (BFI) are characteristics of the sea states that can only be estimated for individual wave systems \citep{Trulsen-2015-JGRO-120-7113,StoleHentschel-2018-PF-30-067102}.
The wave steepness definition employed here is
$\varepsilon=k_p a_c$ where $a_c = H_s/\sqrt{8}$ is a characteristic amplitude
and where $H_s$ is the significant wave height defined as four times the
standard deviation of the surface elevation.
We only consider the BFI of the windsea system $\mathrm{BFI}_W =\varepsilon_W\omega_{p}/\Delta_W$, where $\Delta_W$ is the 'half width at half peak'.
Tab.~\ref{tab:wave_pars} summarizes the characteristics of the wave system employed in this study.

\begin{table}
	\caption{Wave parameters for laboratory experiments and simulations.}
	\label{tab:wave_pars}
	\large
	\begin{tabular}{l c c  }
		& Laboratory & Simulation\\
		\hline\hline
		$T_{p_W}$ [s] & 0.51 &  0.72\\
		$\lambda_{p_W}$ [m] & 0.40 & 0.80 \\
		$T_{p_S}$ [s] & 0.97 & 1.37\\
		$\lambda_{p_S}$  [m]  &1.45 & 2.9\\
		$H_{s_W}$ [cm] & 1.2 & 2.3\\
		$H_{s_S}$ [cm] &  \{0.06, 0.11, 0.17, 0.22\} & \{0.10,0.50,0.90,1.8\}\\
		$\omega_m ~[\omega_p]$  & 0.685 & 0.750\\
		$\varepsilon_W$ ($x=23\lambda_p$) & 0.069  & 0.062 \\
		$\varepsilon_S$ ($x=23\lambda_p$) & $\leq0.003$ & $\leq 0.009$\\
		$\mathrm{BFI}_W $ & $\leq 0.06$ & $\leq 0.25$\\
	\end{tabular}
\end{table}

\begin{figure}[h]
	\centering
	\subfloat[][Laboratory]{\includegraphics[width=.49\textwidth]{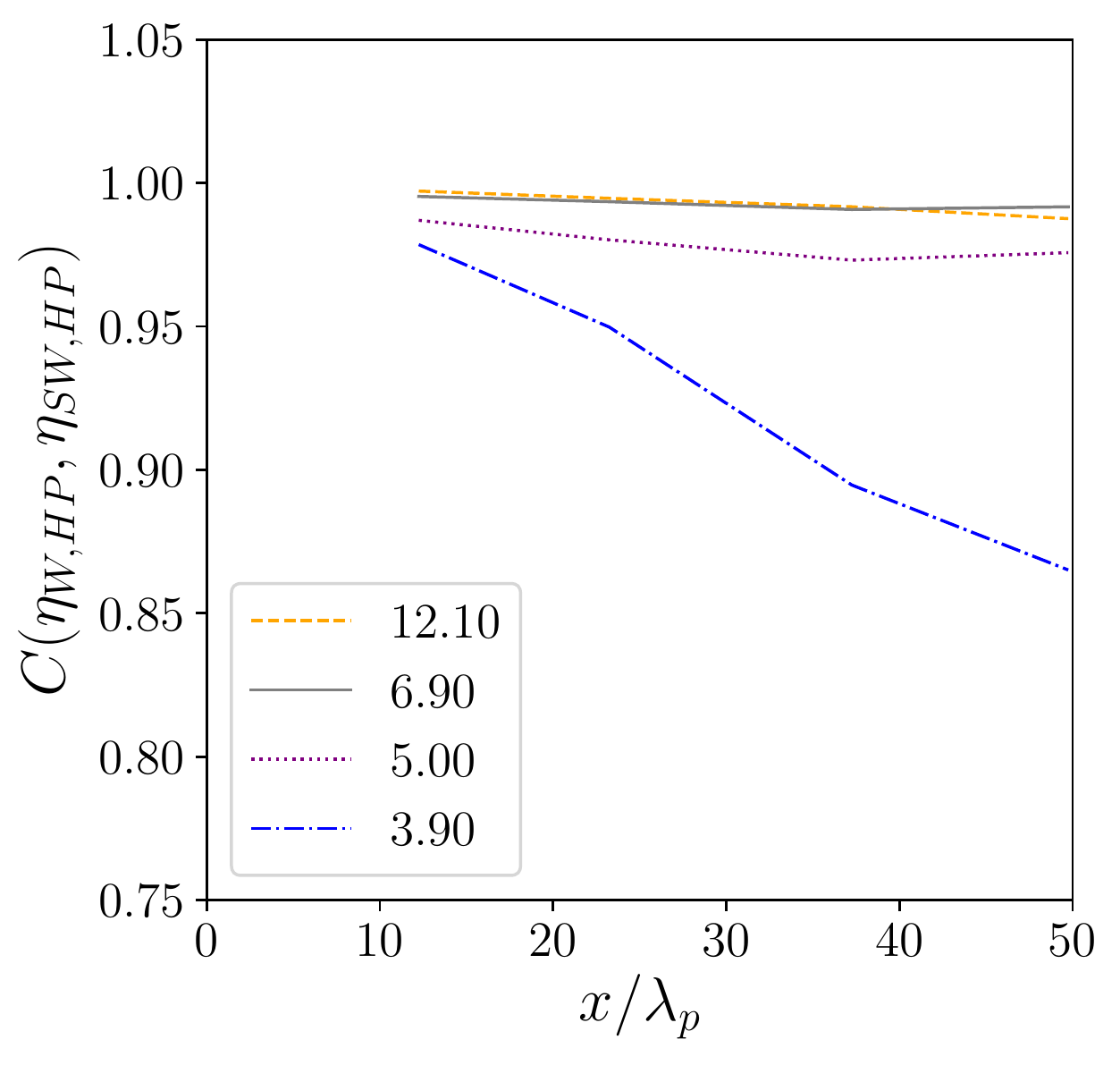}}
	\subfloat[][Simulation]{\includegraphics[width=.49\textwidth]{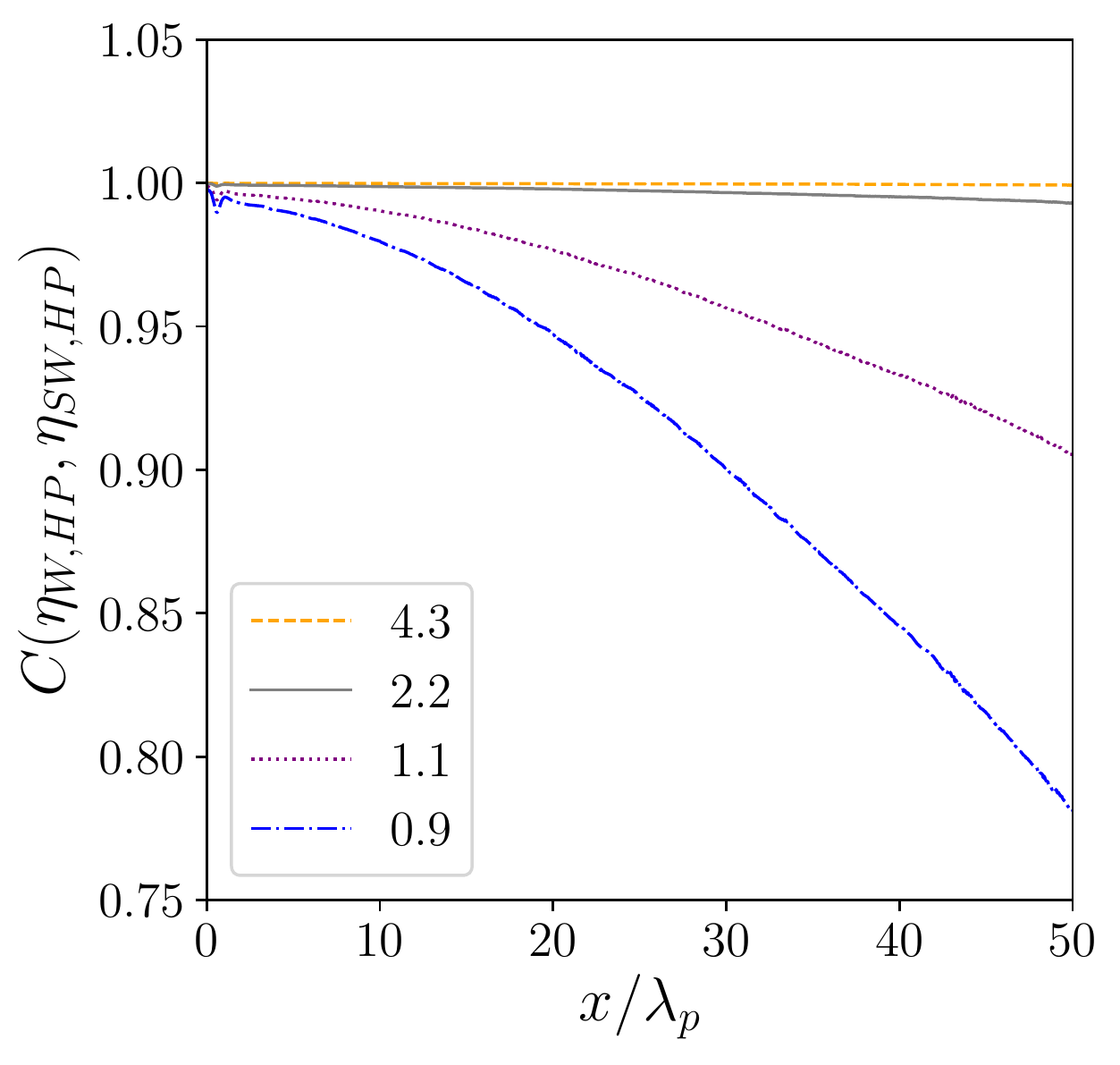}}
	\caption{The correlation coefficient between highpass filtered windsea and highpass filtered mixed cases with the indicated SSER.}
	\label{fig:evolution}
\end{figure}
As a first estimate of the degree of interaction between swell and windsea we calculate the correlation coefficient of pure windsea waves and the windsea part of a combined windsea and swell sea state,
\begin{align}
C(\eta_{\mathrm{W,HP}},\eta_{\mathrm{SW,HP}}) = \frac{\mathrm{Cov}(\eta_{\mathrm{W,HP}},\eta_{\mathrm{SW,HP}})}{\sigma_{\mathrm{W,HP}} \sigma_{\mathrm{SW,HP}}}=\frac{E[\eta_{\mathrm{W,HP}}\eta_{\mathrm{SW,HP}}]}{\sigma_{\mathrm{W,HP}} \sigma_{\mathrm{SW,HP}}},
\end{align}
where $\eta_\mathrm{W,HP}$ and $\eta_\mathrm{SW,HP}$ are the surface elevations of pure windsea and combined swell and windsea both filtered by a high pass filter. The corresponding standard deviations are denoted $\sigma_\mathrm{W,HP}$ and $\sigma_\mathrm{SW,HP}$.
In the case of the simulations, the correlation coefficient shown in Figure~\ref{fig:evolution}b was averaged over all simulations.
Close to the wavemaker, the pure windsea and the windsea parts of the mixed seas are practically identical for all cases.  In the development along the tank the correlation coefficient decays depending on the strength of the swell. Close to the damping beach, the correlation coefficient decays to a level of 0.85 for the laboratory and 0.75 for the simulation for the strongest swell. These observations suggest that the swell has only marginal influence on the phases of the windsea waves over the given evolution length.

\section{Kurtosis and skewness for different sea states}

\begin{figure}
	\centering
	\subfloat[][Laboratory]{\includegraphics[width=0.49\textwidth]{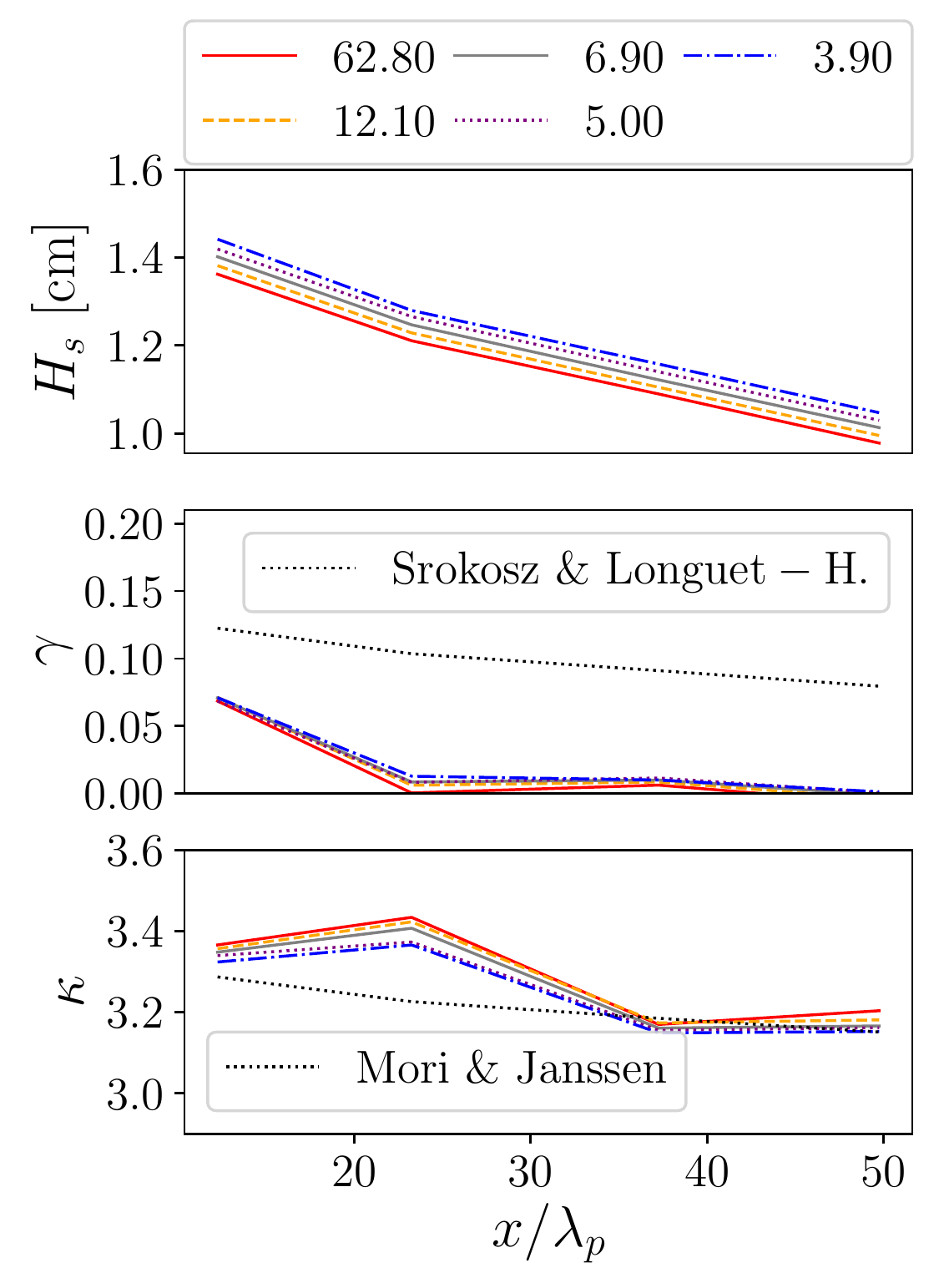}}
	\subfloat[][Simulation]{\includegraphics[width=0.49\textwidth]{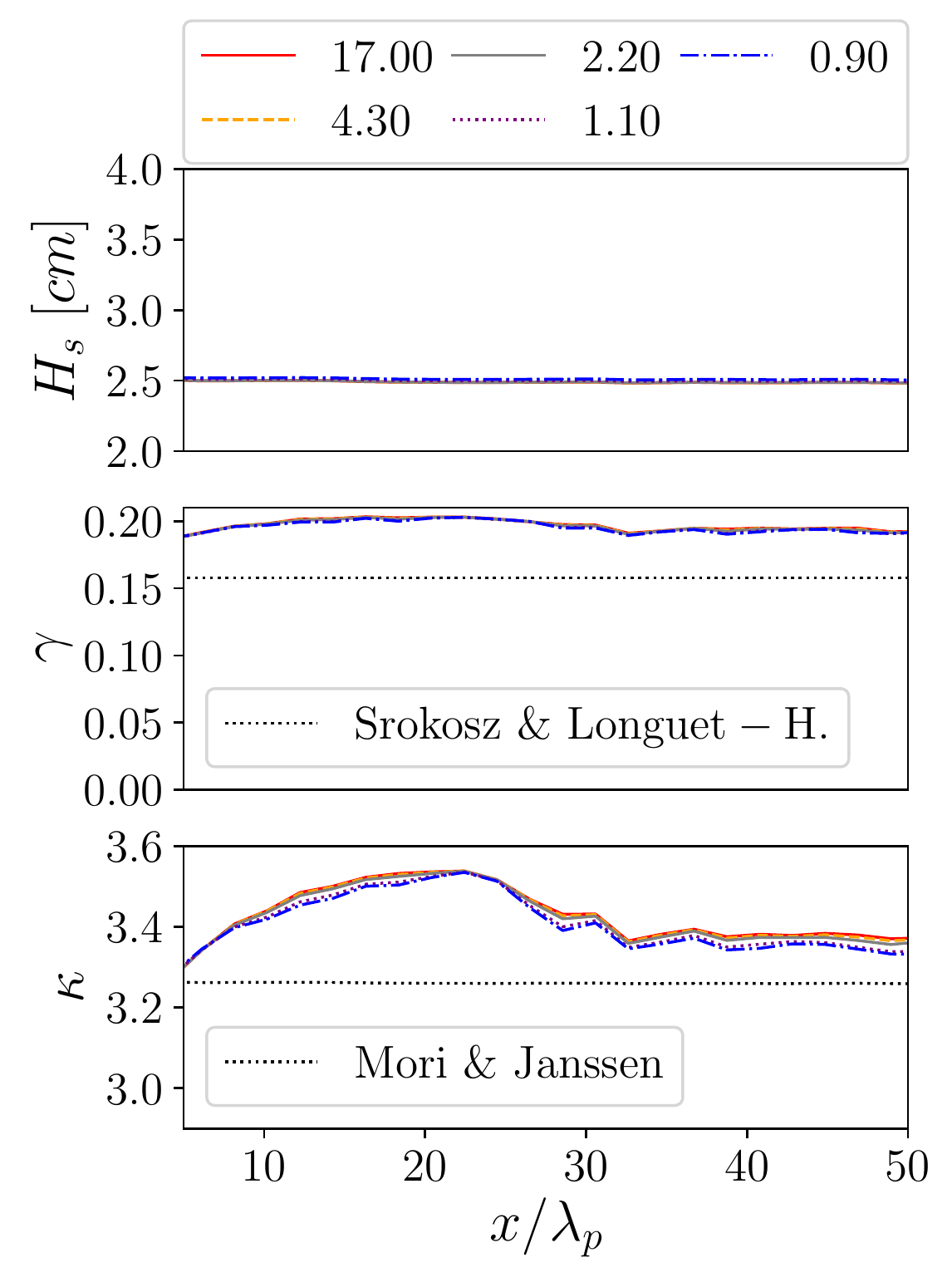}}
	\caption{
		Significant wave height, skewness and kurtosos for the highpass
		filtered waves, labeled by the SSER coefficient.
		With a confidence of 80\%, the deviation from the mean is [-0.07, 0.02] for skewness and [-0.1,0.4] for kurtosis in the laboratory. The corresponding intervals for the simulations are [-0.1,0.2] and [-0.5,0.9]. For comparison we added theoretical results  by \citet{Srokosz-1986-JFM-164-487,Mori-2006-JPO-36-1471} for longcrested waves of narrow bandwidth.}
	\label{fig:res}
\end{figure}

%\begin{figure}
%\centering
%\includegraphics[width=0.49\textwidth]{Spectra_at_374}
%\includegraphics[width=0.49\textwidth]{Hs_kurt_skew_sim_err_HP}
%\caption{Spectra defining the different cases (left) and corresponing $H_s$ kurtosis skewness along the tank for simulations with 80\% confidence intervals (right). The legend gives %the SSER coefficient as defined \citet{GuedesSoares-1984-OE-11-185}. The red solid line corresponds to pure windsea. All spectra are scaled by the maximum of the windsea spectrum.}
%\label{fig:Sim_along_tank}
%\end{figure}
Figure~\ref{fig:res} presents the laboratory measurements and the simulation results in terms of $H_s$, skewness and kurtosis for the highpass filtered windsea waves. In the case of the laboratory, the local skewness, $\gamma$, and the local kurtosis, $\kappa$, were computed  for each fixed position along the tank. The damping beach in the laboratory is too short for the longest waves leading to seiche modes. These have been filtered by a high pass filtered for the entire data analysis. 
For the simulations, we have employed a lower time resolution of $\Delta t = 0.14$ s and the part of the simulation time subject to analysis is limited to 77 $T_p$. To ensure sufficient surface elevation measurements for adequate estimates or kurtosis and skewness, 32 neighboring grid points were considered as one evaluation point in space, and 10 simulations were merged; thus, the mean values and the confidence intervals are based on 50 simulations arranged into five groups.
The strong decay of $H_s$ in the laboratory is attributed to dissipation that has a strong effect on the experiment due to the short peak wavelength of the windsea spectrum. This reduction in wave steepness affects the kurtosis and the skewness, as reflected in the reference curves based on \citet{Srokosz-1986-JFM-164-487} and \citet{Mori-2006-JPO-36-1471}. For the simulations, the significant wave height is constant along the tank. 
Overall, the difference in skewness and kurtosis for the different spectra is smaller than the variation of the values.

\begin{figure}
	\centering
	\subfloat[][Laboratory]{\includegraphics[width=0.49\textwidth]{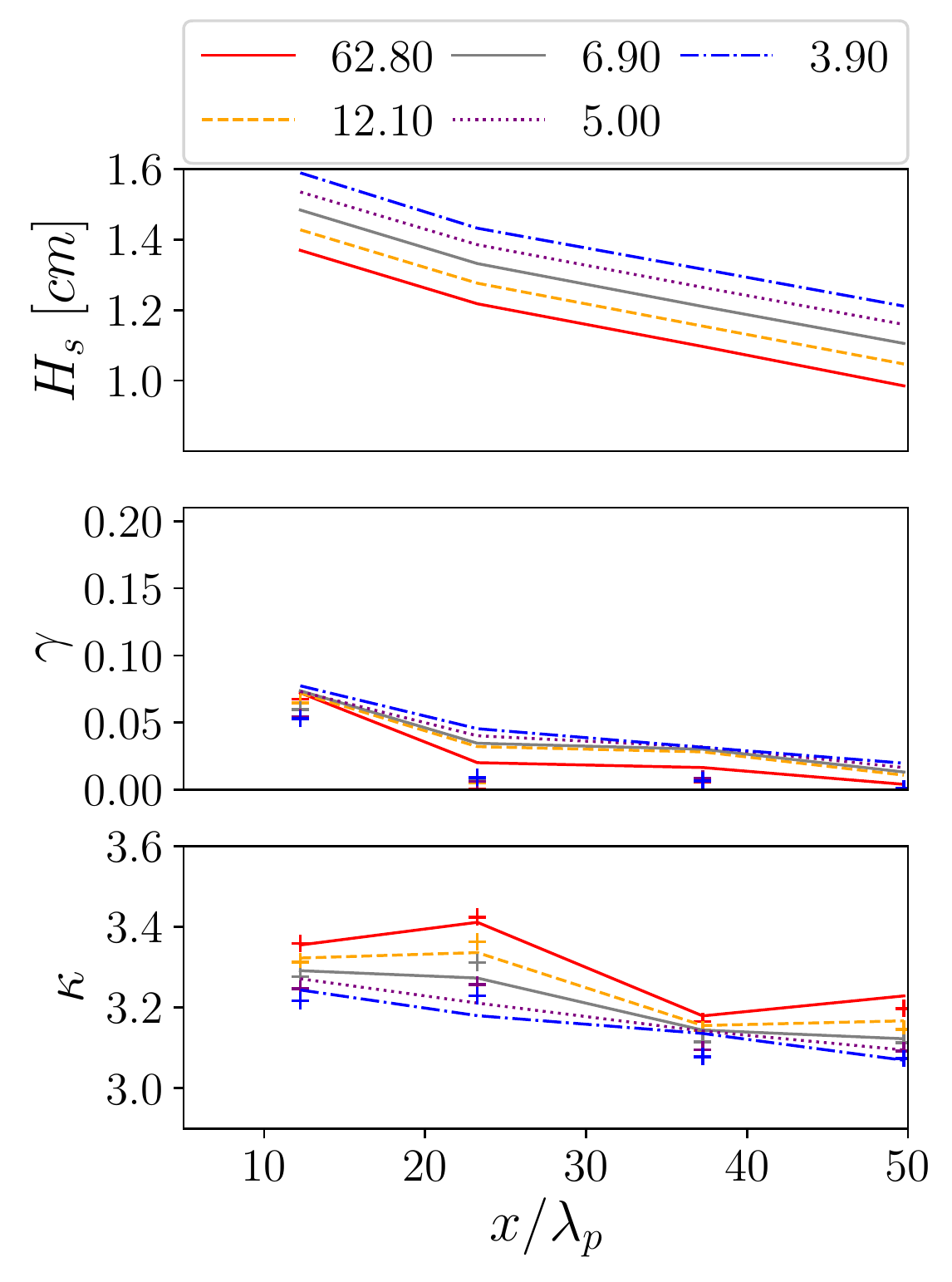}}
	\subfloat[][Simulation]{\includegraphics[width=0.49\textwidth]{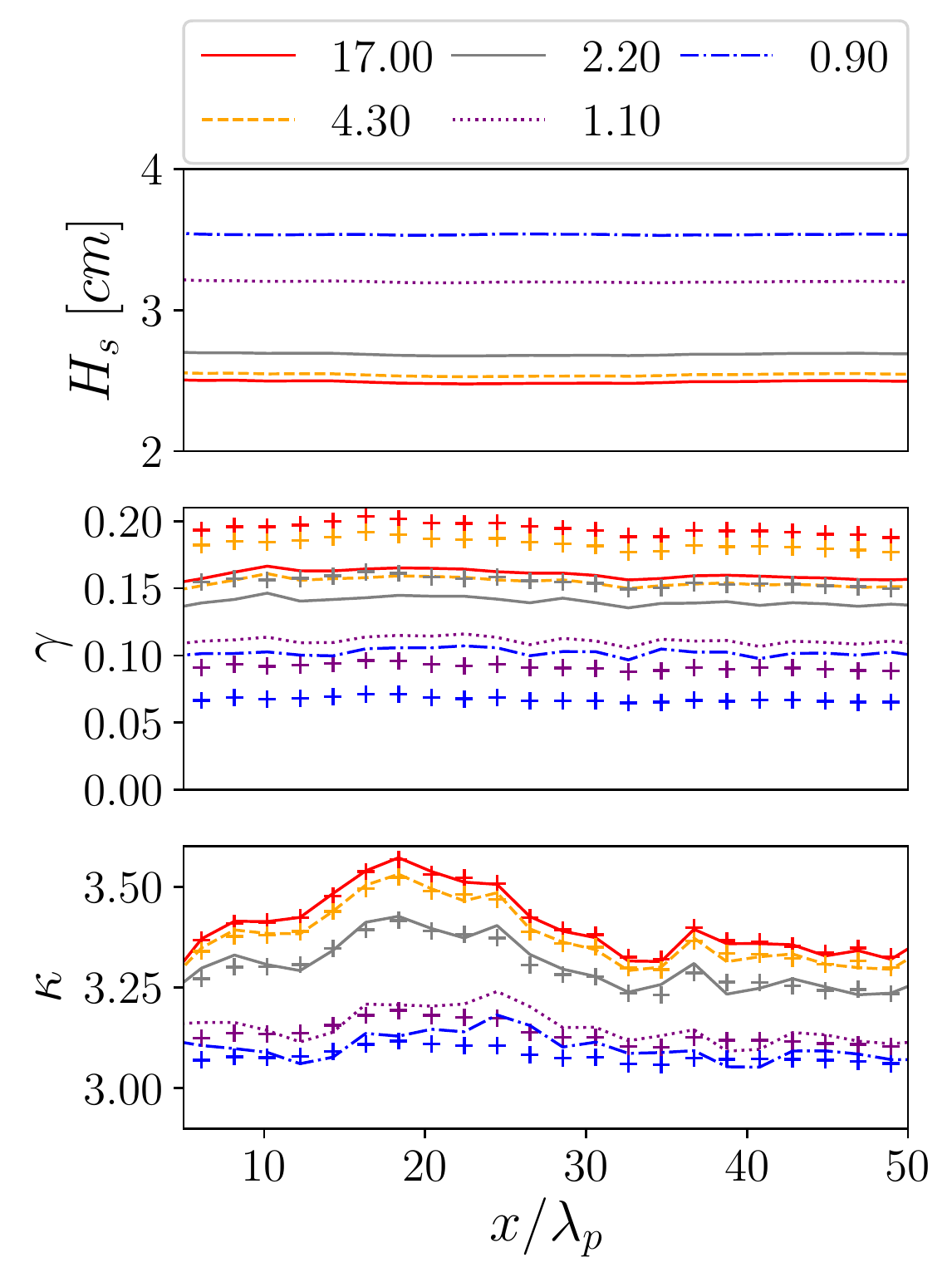}}
	\caption{Significant wave height, skewness and kurtosis for
		the combined windsea and swell wave field (lines). Estimates based on Eq.~\ref{eq:skewest} and~\ref{eq:kurtest} are marked by '+' with the same 
		colors as the corresponding full sea states.  With a confidence of 80\%, the deviation from the mean is [-0.007, 0.05] for skewness and [-0.06,0.4] for kurtosis. The corresponding intervals for the simulations are [-0.02,0.06] and [-0.09,0.32].}
	\label{fig:skew_kurt_unfiltered}
\end{figure}

We also analyze the local skewness and the local kurtosis for the waves of the entire spectra (Figure~\ref{fig:skew_kurt_unfiltered}). Both the laboratory and the simulation results show that the SSER is correlated to the kurtosis of combined seas. The same is shown for the skewness in the simulations. In the laboratory, the skewness behaves differently, however the variation between the cases is small.

From the definition of skewness and kurtosis, we calculate reference values for skewness and kurtosis of combined sea states based on the assumption that the two systems are independent. From the analysis of the correlation coefficient, we anticipate that this may be a good approximation. We find the formulas
\begin{equation}
\gamma_{\mathrm{SW}} = \frac{\gamma_{\mathrm{S}}\sigma_{\mathrm{S}}^3 + \gamma_{\mathrm{W}}\sigma_{\mathrm{W}}^3}{(\sigma_{\mathrm{S}}^2 + \sigma_{\mathrm{W}}^2)^{3/2}}
\label{eq:skewest}
\end{equation}
and
\begin{equation}
\kappa_{\mathrm{SW}} = \frac{\kappa_{\mathrm{S}}\sigma_{\mathrm{S}}^4 + 6\sigma_{\mathrm{S}}^2\sigma_{\mathrm{W}}^2 + \kappa_{\mathrm{W}}\sigma_{\mathrm{W}}^4}{(\sigma_{\mathrm{S}}^2 + \sigma_{\mathrm{W}}^2)^2}
\label{eq:kurtest}
\end{equation}
with $\{\sigma_\mathrm{S}^2,\gamma_\mathrm{S},\kappa_\mathrm{S}\}$ the variance, skewness and kurtosis of the swell partition, and $\{\sigma_\mathrm{W}^2,\gamma_\mathrm{W},\kappa_\mathrm{W}\}$ the variance, skewness and kurtosis of the windsea partition.
Figure~\ref{fig:skew_kurt_unfiltered} shows that the analysis of the entire wave system differs clearly from the analysis of the windsea system (Figure~\ref{fig:res}). The estimates in Eq.~\ref{eq:skewest} and~\ref{eq:kurtest} are similar but not equal to the calculated values.

The kurtosis of the windsea waves is generally high in the presented analysis which is expected for longcrested waves. According to \citet{Gramstad-2007-JFM-582-463}, wave crests longer than 
ten times the wavelength are associated with very high kurtosis values, even for large spectral bandwidths that are not typically associated with modulational instability 
according to a low Benjamin-Feir index \citep{Janssen-2003-JPO-33-863}.

\section{Exceedance probability of wave envelope}

\begin{figure}[h]
	\centering
	\includegraphics[width=.49\textwidth]{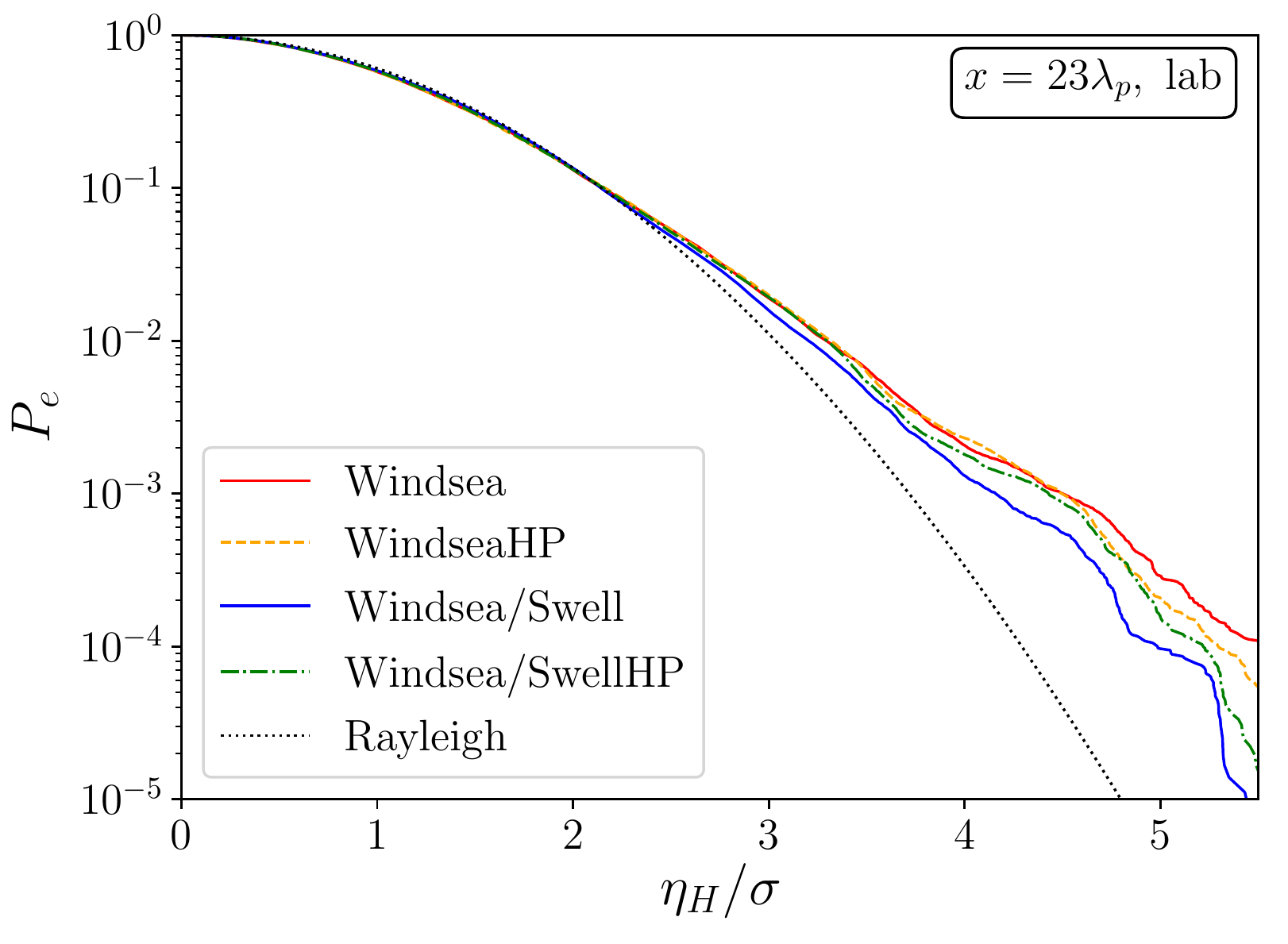}
	\includegraphics[width=.49\textwidth]{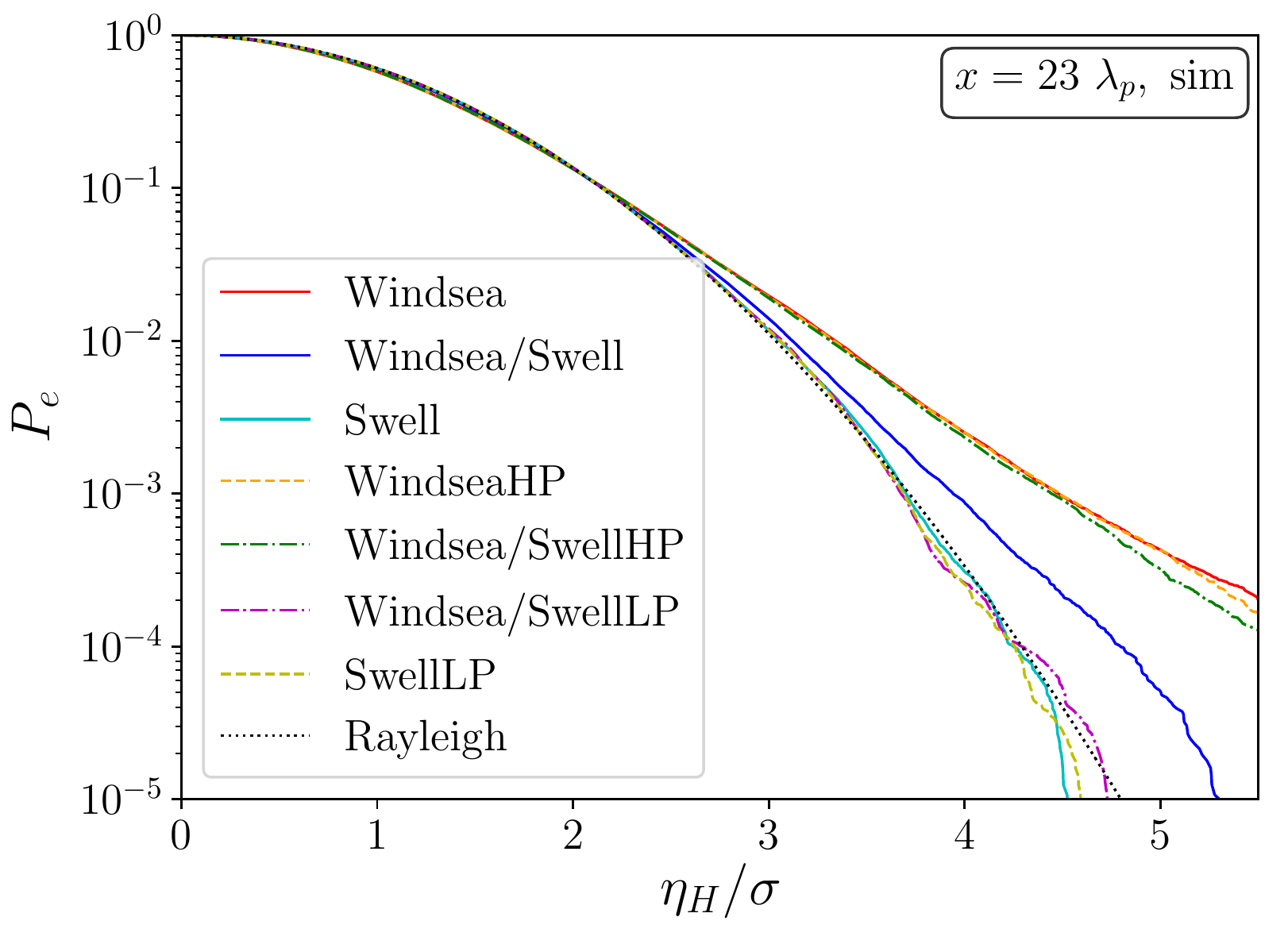}	
	\caption{Exceedance probability of Hilbert envelopes scaled by the standard deviation of the underlying surface elevation at one position. The mixed sea  is based
		on the maximum swell for laboratory (SSER=3.9) and simulation (SSER=0.9), respectively. The windsea has a SSER of 62.8 for the laboratory and 17.0 for the simulation.}
	\label{fig:exceed_hilbert}
\end{figure}
We study the exceedance probability of the Hilbert envelope to retrieve insight in the amount
of extreme events in the different cases. The Hilbert wave envelope was computed according to
\begin{equation}
\eta_H = \sqrt{\eta^2 + \tilde{\eta}^2}
\end{equation}
where $\tilde{\eta}$ is the Hilbert transform of the surface elevation $\eta$.
The envelopes were scaled by the standard deviation of the corresponding surface elevation.
For reference, we compare the probability distributions
of the different cases with the Rayleigh distribution, which is the distribution of the Hilbert envelope
of a Gaussian process.

In the presentation of the exceedance curves in this and the next chapter, we focus on pure windsea and the mixed sea with the strongest swell for laboratory and simulation. Pure swell simulations were added for comparison.
The different sea states are analysed both filtered and unfiltered.
The results in Figure~\ref{fig:exceed_hilbert} are clearer for the simulations than for the laboratory experiments. 
The swell envelopes of all configurations coincide with the Rayleigh distribution in good correspondence with the fact that the swell is essentially linear. 
While the envelopes of the complete wave fields differ, their windsea contributions are similar and deviate clearly from Gaussian behavior.

\section{Exceedance probability of crest height}

\begin{figure}[h]
	\centering
	\includegraphics[width=.49\textwidth]{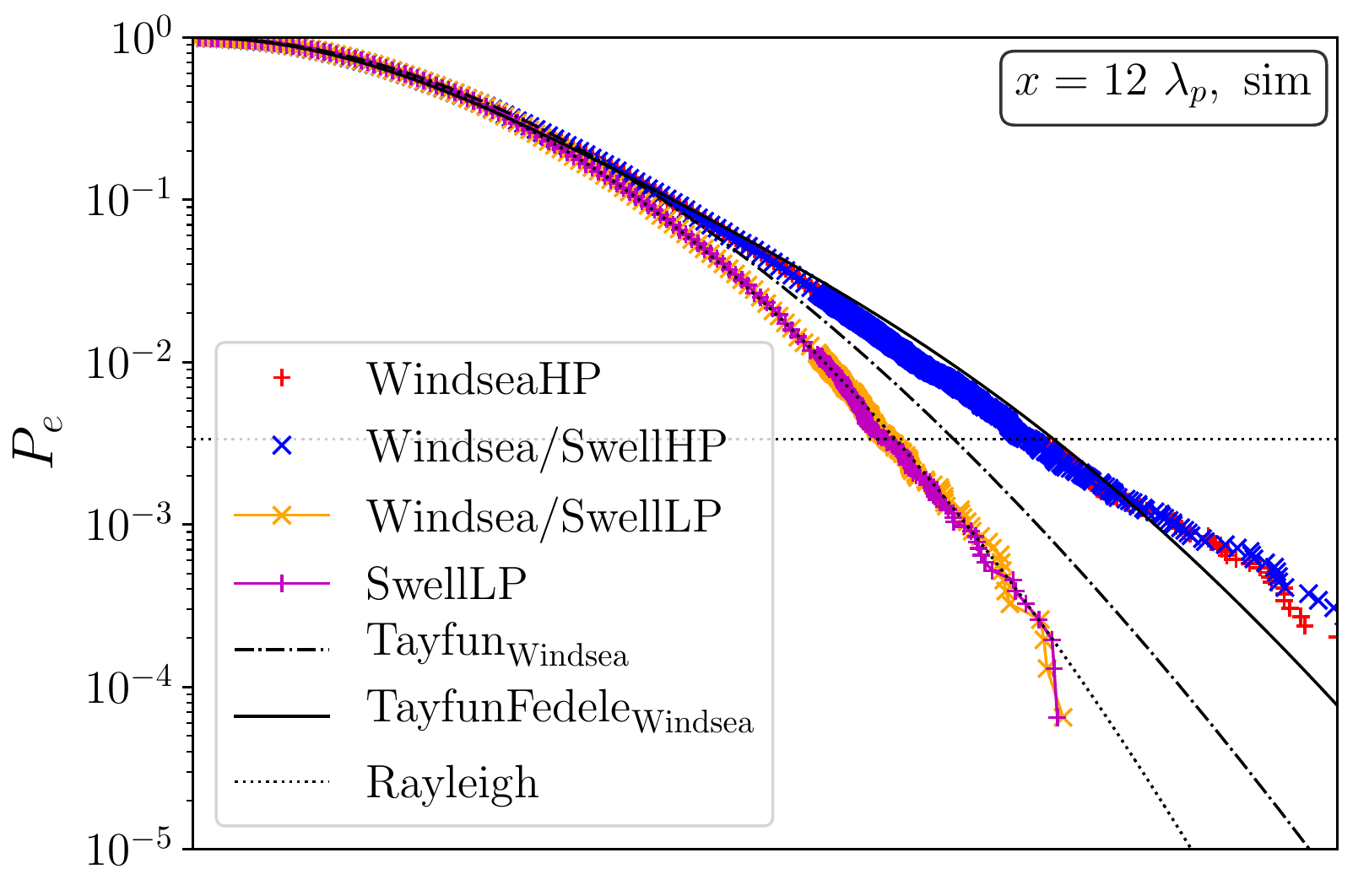}
	\includegraphics[width=.49\textwidth]{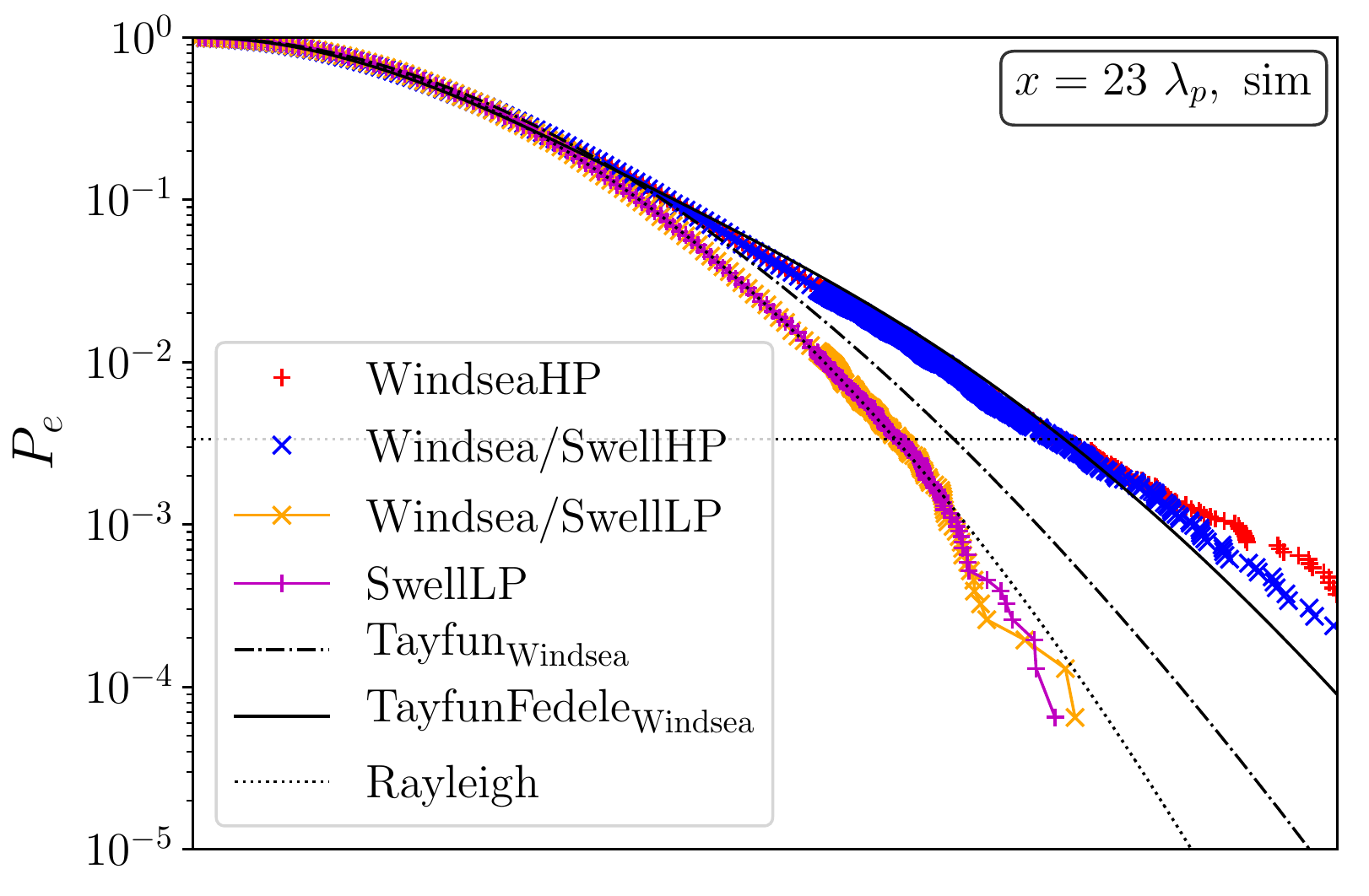}\\
	\includegraphics[width=.49\textwidth]{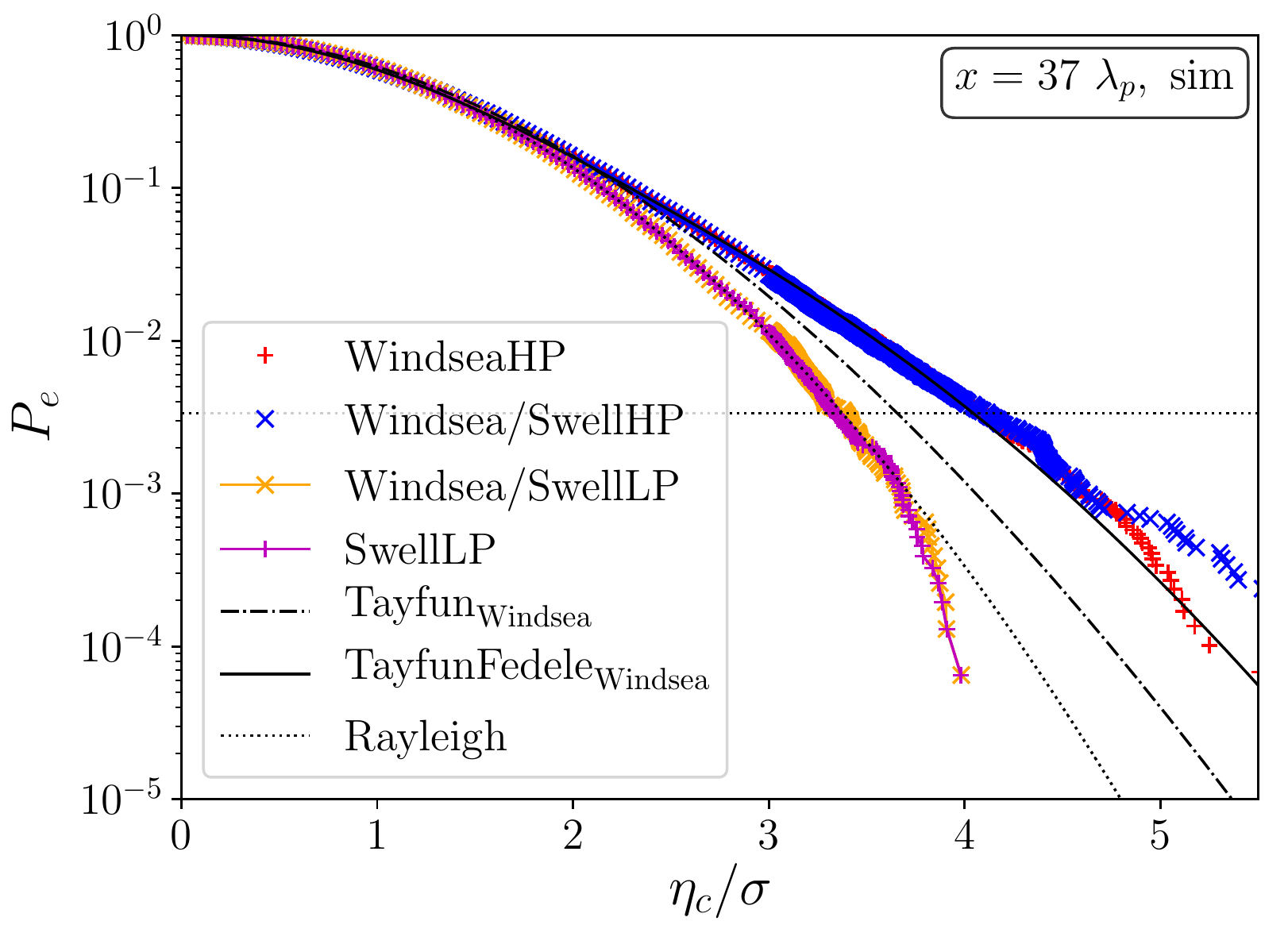}
	\includegraphics[width=.49\textwidth]{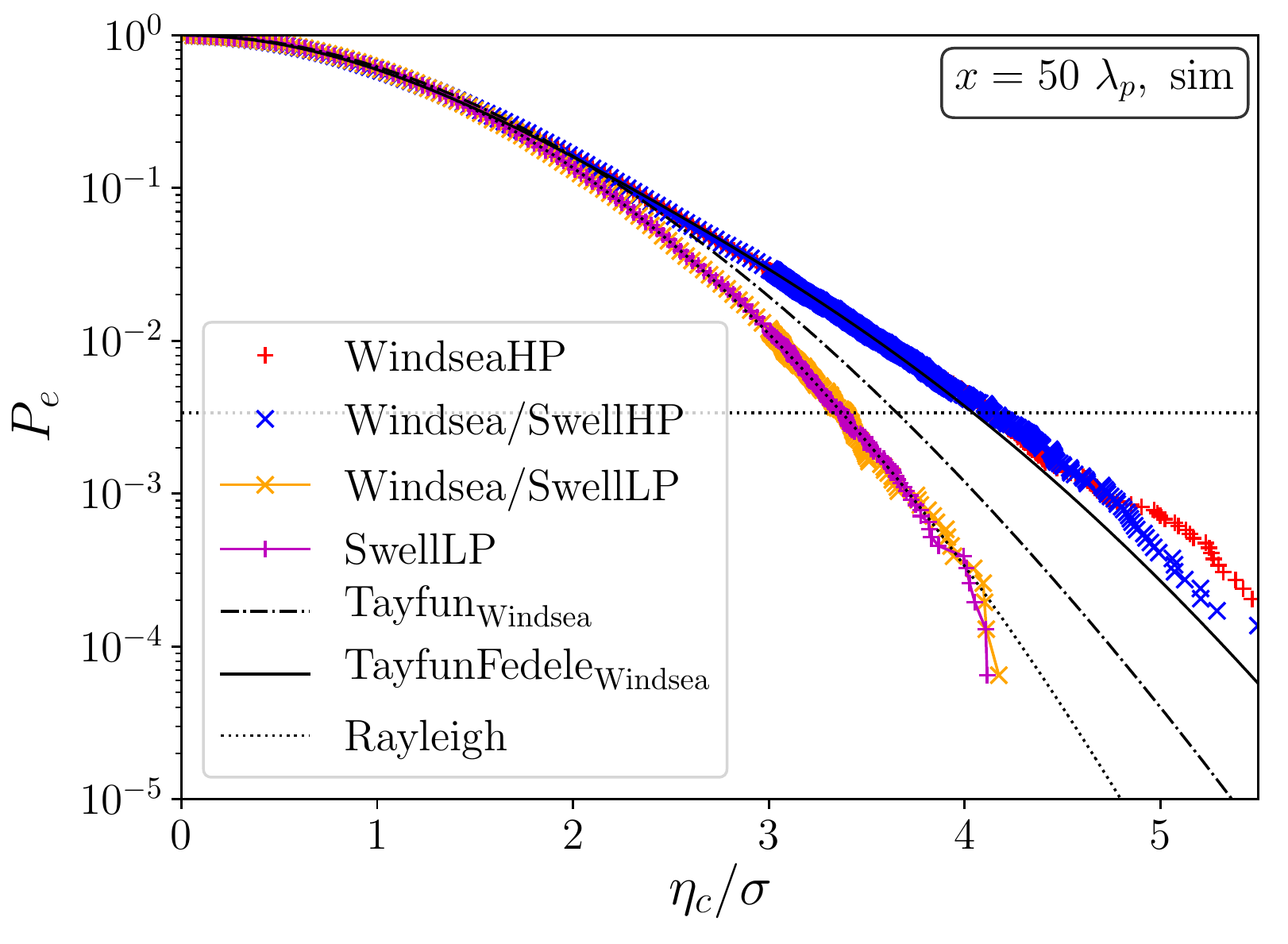}
	
	\caption{Exceedance probability of crest heights scaled by the standard deviation of the corresponding surface elevation at four positions along the numerical wave tank. SSER values are 17 and 0.9 for windsea and mixed seas respectively.}
	\label{fig:exceed}
\end{figure}

Finally, we investigate the exceedance probability of the crest height, $\eta_c$ scaled by the standard deviation, $\sigma$ of the corresponding surface elevation (see Figure~\ref{fig:exceed}).
We only consider the exceedance probability of the crests for the simulations since the number of measured crests in the laboratory was too small for meaningful analysis. 
For better results on extreme cases, the number of simulations was increased to 180, compared to that of Tab.~\ref{tab:sim_pars}. 
The simulated data was interpolated by truncated Fourier series expansion in time before determining the crest height as maximum between two zero crossings. 
Below the horizontal dotted lines the number of crests is lower than one hundred which is considered too few for statistical significance.

The exceedance probability is characterized by local variation, similar to what
we previously found for the skewness and the kurtosis.  
To
illustrate the amount of local variation, we show the
exceedance probability at four locations along the numerical wavetank (see Figure~\ref{fig:exceed}).

When comparing the waves after partitioning into swell and windsea, the statistics are approximately identical irrespective if the other wave system is present or not.

As reference distributions, we consider Rayleigh, Tayfun~\citep{Tayfun-1980-JGR-85-1548} and Tayfun-Fedele~\citep{Fedele-2016-SR-6-27715,Tayfun-2007-OE-34-1631}.
The wave crests of narrow banded linear waves are Rayleigh distributed. Although the simulated swell is not particularly narrow-banded, its crests are found to be Rayleigh distributed. The Tayfun distribution, here calculated based on the standard deviation of the windsea waves, is the reference distribution for narrow banded second order wave crests. The windsea distributions clearly exceed Tayfun even though they are clearly not narrow-banded.  Hence we anticipate that nonlinearities of third order must play an important role for the windsea waves. As an approximation to a third order distribution of crest heights we employ the Tayfun-Fedele distribution. It is a continuation of the second-order Tayfun distribution by the Gram-Charlier series up to the fourth cumulant~\citep{Fedele-2016-SR-6-27715,Tayfun-2007-OE-34-1631}. The distribution changes its form with the wave steepness by its dependence on the skewness and the occurrence of extreme values by the incorporation of the kurtosis. The Tayfun-Fedele distribution shows good agreement with the simulated windsea exceedance probability of crest heights.

\begin{figure}[h]
	\centering
	\includegraphics[width=.49\textwidth]{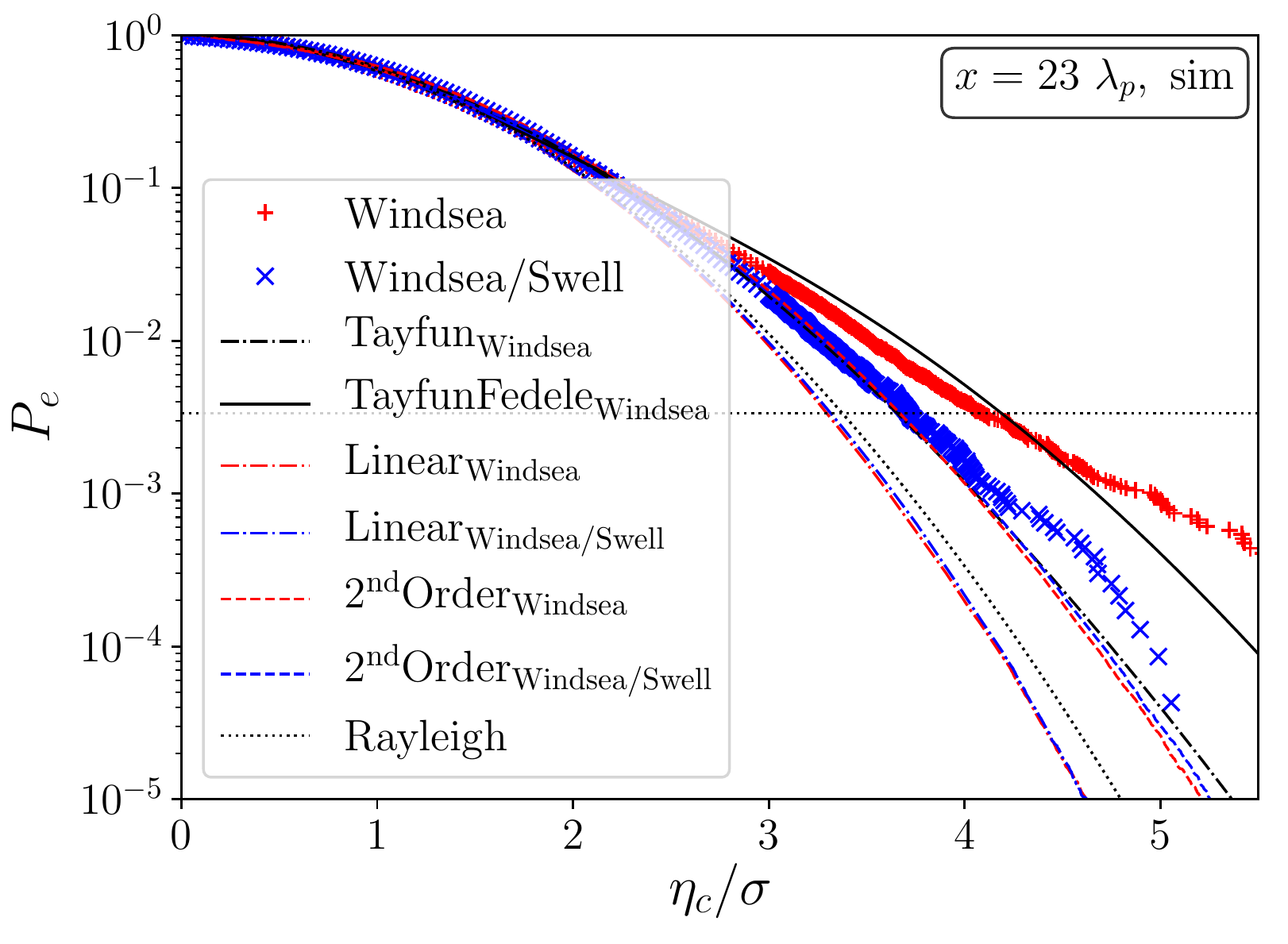}	
	\caption{Exceedance probabilities of crest heights of unfiltered sea states. SSER values are 17 and 0.9 for windsea and mixed seas respectively.}
	\label{fig:revisited}
\end{figure}

Figure~\ref{fig:revisited} reveals that the combined  swell and windsea results in an exceedance probability for crest heights between that of pure swell and that of pure windsea that is coincidentally similar to that of narrow-banded waves of second order. Simulations of linear theory and second order theory \citep{Forristall-2000-JPO-30-1931} were added.

%     KT
% Though the spectra are clearly not narrow banded, especially when windsea and swell are combined, their deviation form the corresponding distribution for narrow bandwidth are surprisingly little.

\section{Discussion}
Our results show that combined sea states require a different analysis than single wave systems.
A unified steepness cannot be defined for the combination of multiple wave systems \citep{Naciri-1992-JFM-235-415, Naciri-1993-PFA-5-1869, Gramstad-2010-JFM-650-57, Gramstad-2011-PF-23-062102, StoleHentschel-2018-PF-30-067102, Trulsen-2015-JGRO-120-7113}.
We have shown that the partitioning of the spectrum is not only necessary for estimating the various partial steepnesses, but also for better understanding of the statistics of extreme waves.

%We observe that the deviation between the statistics of the windsea cases is smaller for the simulations than for the laboratory. 
%The Torsethaugen spectrum applied in the laboratory produces a swell with a long tail which adds considerable energy to the windsea part of the spectrum.
%This additional energy differs for all cases which manifests in the statistics of the windsea part of the spectrum.
%The problem was not present in the simulation where the swell only affects the windsea spectrum for frequencies slightly above the splitting frequency $\omega_m$. 
%Even with a stronger variation of the strength of the swell, the windsea waves remain essentially unaffected. 

We observe that the results are more distinct for the simulations than for the laboratory, which we attribute mainly to the different spectra that were applied. 
The windsea parts of the Torsethaugen spectra differ considerably for minor changes of the swell part, while the windsea part of Ochi \& Hubble spectra are very similar, 
even for very different swell contributions.

%Eq.~\ref{eq:skewest} and~\ref{eq:kurtest} show that skewness and kurtosis of the combined system of a nonlinear windsea and a dominant linear swell can be much smaller than the skewness and kurtosis of the windsea alone.  
As formulated in Eq.~\ref{eq:skewest} and~\ref{eq:kurtest}, skewness and kurtosis for mixed sea are weighted combinations of skewness and kurtosis of the separated systems. Therefore, the combination of a Gaussian and a strongly non-Gaussian system of equivalent $H_s$ will appear weakly non-Gaussian.
%A similar behavior seems to occur for the exceedance probabilities.  On the other hand, we have evidence that the swell does not affect the windsea much.  In this case we believe that extreme wave statistics for the combined sea may underestimate how dangerous the combined sea state really is.
A similar behavior seems to occur for the exceedance probabilities.  
In our study, the swell affects the windsea so little that the two systems may be regarded as independent. In this case we believe that extreme wave statistics for the combined sea may underestimate how dangerous the combined sea state really is.

\section{Conclusion}

We have shown that extreme wave statistics of windsea can be nearly unaffected by the presence of a following swell.  However, this only becomes apparent if the two sea states are analysed separately.
Analysis of the entire wave system as a whole gives the impression that the mixed sea has milder extreme wave statistics than the pure windsea, especially when the nonlinearity in the two systems clearly differs.  Therefore we believe that mixed sea states should be partitioned, and analysis be performed on each partition, in order to fully comprehend the combined extreme wave statistics.

%     KT
%     The present work demonstrates that windsea waves have approximately identical exceedance probability distriubtions irrespective if a linear swell is present or not. 
% However, the result only becomes apparent if the sea states are analysed separately.
% The analysis of the entire wavesystem as a whole gives the impression that the mixed sea is a less severe sea state compared to the pure windsea, especially when the degree of non-linearity in the two systems clearly differs.

% We believe the findings may prove (???) relevant for the analysis of a large variety of sea states, without the restrictions to following seas and long-crested waves.

\section*{Acknowledgements}
	We thank Jos\'e Carlos Nieto Borge for important discussions, especially for sharing his previous experience from the department of Clima Mar\'itimo, at Puertos del Estado in Madrid, Spain, where they had fully realized the importance of partitioning the wave spectra as a tool for wave data analysis.
	This research has been funded by the Research Council of Norway (RCN) through projects RCN 214556 and RCN 256466.

\bibliographystyle{spbasic}

\bibliography{../biblio}

\end{document}